\documentclass[apj]{emulateapj}

\usepackage{graphicx}
\usepackage{apjfonts}

\newcommand{\kms}{~km~s$^{-1}$}
\newcommand{\msun}{{M$_{\odot}$}}
\newcommand{\dg}{$^{\circ}$}
\newcommand{\drho}{$\rho_{shock}/\rho_0$}
\bibpunct{(}{)}{,}{a}{}{}

\shorttitle{Filaments, Bubbles, and Weak Shocks in M87} 
\shortauthors{Forman et al.}

\begin{document}

\title{Filaments, Bubbles, and  Weak
   Shocks in the Gaseous Atmosphere of M87}

\author{W.~Forman\altaffilmark{1}, 
C.~Jones\altaffilmark{1},
E.~Churazov\altaffilmark{2,3},
M.~Markevitch\altaffilmark{1},
P.~Nulsen\altaffilmark{1},
A.~Vikhlinin\altaffilmark{1,2},
M.~Begelman\altaffilmark{4},
H.~B\"ohringer\altaffilmark{5},
J.~Eilek\altaffilmark{6},
S.~Heinz\altaffilmark{7},
R.~Kraft\altaffilmark{1},
F.~Owen\altaffilmark{8}
M.~Pahre\altaffilmark{1}\\
{\bf Short Title: Filaments, Bubbles, and Weak Shocks in M87} }

\altaffiltext{1}{Smithsonian Astrophysical Observatory,
Harvard-Smithsonian Center for Astrophysics, 60 Garden St., Cambridge,
MA 02138; wforman@gmail.com}
\altaffiltext{2}{Space Research Institute (IKI), Profsoyuznaya 84/32,
Moscow 117810, Russia}
\altaffiltext{3}{MPI f\"{u}r Astrophysik, Karl-Schwarzschild-Strasse
1, 85740 Garching, Germany}
\altaffiltext{4}{JILA, 440 UCB, University of Colorado,
Boulder, CO 80309}
\altaffiltext{5}{MPI f\"{u}r Extraterrestrischephysik,
Giessenbachstrae, 85748 Garching, Germany}
\altaffiltext{6}{New Mexico Tech., Socorro, NM 87801}
\altaffiltext{7}{University of Wisconsin, Madison, WI}
\altaffiltext{8}{National Radio Astronomy Observatory, Socorro, NM
87801}

\begin{abstract}

We present the first results from a 500~ksec Chandra ACIS-I
observation of M87.  At soft energies (0.5-1.0 keV), we detect
filamentary structures associated with the eastern and southwestern
X-ray and radio arms.  Many filaments are spatially resolved with
widths of $\sim300$~pc. This filamentary structure is particularly
striking in the eastern arm where we suggest the filaments are outer
edges of a series of plasma-filled, buoyant bubbles whose ages differ
by $\sim6\times10^6$~years. These X-ray structures may be influenced
by magnetic filamentation.  At hard energies (3.5-7.5 keV), we detect
a nearly circular ring of outer radius $2.8'$ (13 kpc) which provides
an unambiguous signature of a weak shock, driven by an outburst from
the SMBH. The density rise in the shock is \drho$\approx1.3$ (Mach
number, $M\approx1.2$).  The observed spectral hardening in the ring
corresponds to a temperature rise $T_{shock} / T_0 \approx 1.2$, or
$M\approx1.2$, in agreement with the Mach number derived independently
from the gas density. Thus, for the first time, we detect gas
temperature and density jumps associated with a classical shock in the
atmosphere around a supermassive black hole. We also detect two
additional surface brightness edges and pressure enhancements at radii
of $\sim0.6'$ and $\sim1'$. The $\sim0.6'$ feature may be
over-pressurized thermal gas surrounding the relativistic plasma in
the radio cocoon, the ``piston'', produced by the current episode of
AGN activity.  The over-pressurized gas is surrounded by a cool gas
shell. The $\sim1'$ feature may be an additional weak shock from a
secondary outburst. In an earlier episode, the ``piston'' was
responsible for driving the $2.8'$ shock.

\end{abstract}


\section{Introduction}

M87 (NGC4486), the dominant central galaxy in the Virgo cluster, hosts
a $3.2\times10^9$~\msun~ supermassive black hole (SMBH, Harms et
al. 1994, Ford et al. 1994, Macchetto et al. 1997) and its
well-studied jet (e.g., Sparks, Biretta, \& Macchetto 1996, Perlman et
al. 2001, Marshall et al. 2002, Harris et al. 2003). On larger scales
M87 has been the subject of detailed radio observations showing
remarkable structures on scales up to 40 kpc (Owen et al. 2000; see
also Hines et al. 1989). In soft X-rays, M87 is the second brightest
extragalactic source (after the Perseus cluster) and the emission is
dominated by thermal radiation from its $\sim2$ keV gaseous atmosphere
(e.g., Gorenstein et al. 1977, Fabricant \& Gorenstein
1983, B\"ohringer et al. 2001, Matsushita et al. 2002, Belsole et
al. 2001, Molendi 2002).

M87 is a classic example of a ``cooling flow'' cluster (e.g., Fabian
1994).  XMM-Newton and Chandra observations (e.g., David et al. 2001,
Peterson et al. 2003 and references therein) limit the amount of
cooling gas in cluster cores to 10-20\% of that predicted by the
standard cooling flow model and suggest that the average core
temperature does not fall below $\sim30$\% of the gas temperature at
large radii. The dramatic reduction in the amount of cooling gas,
compared to the standard model, requires a considerable energy supply
to compensate for the observed radiative losses.  Thus, instead of
seeking the repository of cold gas, we are now seeking the energy
source to (re)heat the radiating gas.

Suggestions for heating include thermal conduction (e.g., Tucker \&
Rosner 1983, Bertschinger \& Meiksin 1986, Gaetz 1989, David, Hughes
\& Tucker 1992, Zakamska \& Narayan 2003, Voigt \& Fabian 2004) and
cluster mergers (e.g., Fujita et al. 2004, but see B\"ohringer et
al. 2004).   One of the most promising heating mechanisms has been
SMBH outbursts.  Strong shocks are rare and most heating probably
arises from less violent mechanisms to transfer AGN mechanical power
into heating the thermal plasma.  Buoyant bubbles, inflated by SMBH
activity, were first modeled for M87 by Churazov et al. (2001; see
also Reynolds, Heinz \& Begelman 2001, Kaiser \& Binney 2003, DeYoung
2003). These bubbles can be a significant energy source for the
cooling gas and can explain much of the radio and X-ray morphology
(Churazov et al. 2001, Quilis et al. 2001, Churazov et al. 2002,
Br\"uggen et al. 2002).  The energy of jets and buoyant bubbles  
can be transferred to  the radiating gas by via shocks,
gravity waves, or turbulence (e.g., Churazov et al. 2002; Ruszkowski,
Br\"uggen \& Begelman 2004a, b; Begelman 2004; Omma et al. 2004;
Roychowdhury et al. 2004, 2005; Heinz \& Churazov 2005; Mathews,
Faltenbacher \& Brighenti 2006; Binney, Alouani Bibi \& Omma 2007).
From deep Chandra observations of the Perseus cluster, Fabian et
al. (2006) showed that pressure ripples could balance radiative
cooling in the Perseus core, if the viscosity is high.  Dunn et
al. (2005) showed that buoyant bubbles may be a significant heat
source in 70\% of a sample of cooling core clusters (see also Birzan
et al. 2004).

As the nearest cooling core cluster with an active nucleus, M87 is an
ideal system for studying the energy input from the AGN to the hot,
cooling gas. The first suggestion of structure in the
gaseous halo of M87 and its possible relation to the radio structure
was made by Feigelson et al. (1987). Subsequently, with 
ever-improving angular resolution, the relationship has been explored
in more detail (B\"ohringer et al. 1995, Churazov et al. 2001, Belsole
et al. 2001).  In an earlier 40 ksec Chandra observation, Young,
Wilson \& Mundell (2002) reported X-ray cavities 
and edges in the surface brightness profile.  In a
longer ($\sim$ 100 ksec) Chandra observation, Forman et al. (2005)
suggested that the edges or rings of enhanced emission at 13 and 17
kpc were likely shock fronts associated with AGN outbursts that began
1-2 $\times$10$^{7}$ years ago.  From these same observations, Jordan
et al. (2004) identified a population of $\sim$150 LMXBs, 40\% of
which are associated with globular clusters.

We report here on the 500 ksec observation of M87
with the Chandra Observatory. We focus on two major results:
\begin{itemize}
\item the soft emission (0.5-1.0 keV) from M87, especially in the
  eastern and southwestern arms, forms a web of {\em resolved}
  filamentary structures (broader in extent than a point source) of
  width $\sim300$~pc (see Fig.~\ref{fig:soft}).  These features,
  especially those seen in the eastern arm may arise from a series of
  buoyant bubbles at different heights in the atmosphere and different
  stages of evolution.
\item the hard emission (3.5-7.5 keV) shows a ring of emission (see
  Fig.~\ref{fig:hard}) that is nearly circular with an outer radius
  ranging from $2.5'$ to $2.85'$ (11.6 - 13.3 kpc).  This ring of hard
  emission provides an unambiguous signature of a weak shock. Hardness
  ratios (deprojected) show that the gas temperature in the ring rises
  from $\sim2.0$ keV to $\sim2.4$ keV implying a Mach number of $M
  \sim 1.2$ (shock velocity $v = 880 $ \kms~ for a 2 keV thermal
  gas). At the shock, the density jump is 1.33 which yields a Mach
  number of 1.22, consistent with that derived from the temperature
  jump. The age of the outburst that gave rise to the
  shock must be approximately the radius of the shock divided the shock
  velocity, $t_{outburst} \lesssim R_{shock}/v_{shock} = 14$ Myr.
\end{itemize}

In addition to these results, the Chandra image provides a wealth of
information on structures with scales from the jet to large scale
cavities.  We will discuss these other features in more detail in
future papers and focus here on the two new results from the soft and
hard band images from the deep observation.

\section{Observations and Data Processing}

\subsection{Chandra Observations}

The new Chandra observations (OBSIDS 5826, 5827, 5828, 6186, 7210,
7211, and 7212) were taken at a variety of instrument roll angles from
February to November 2005 using the ACIS-I detector (CCDs I0-I3) in
Very Faint (VF) mode to minimize the background. We reprocessed all
observations applying the latest CTI and time dependent gain
calibrations (acisD2000-01-29gain\_ctiN0005.fits). We incorporated
non-uniformity of the quantum efficiency and the time and spatial
dependence of the contamination on the optical blocking filter.  We
performed the usual filtering by grade, excluded bad/hot pixels and
columns, removed cosmic ray ``afterglows'', and applied the VF mode
filtering. We also reprocessed two Faint mode ACIS-S OBSIDS (3717 and
2707) and treated the front (S2) and back (S3) illuminated CCDs
independently.  We compared the images from the Very Faint and Faint
mode observations to verify that no artifacts were introduced near the
bright jet by use of the large $5\times5$ pixel event regions in VF
mode. We examined the data for background flaring and found moderate
flaring in OBSIDS 3717 and 2707 (the back illuminated CCDs only) for
which we excluded approximately half the duration. The total effective
remaining observation time varied over the field, since the
observations were taken at many different roll angles. A typical
effective exposure time is $\sim525$ ksec. The background files (see
Markevitch 2001 for details) were processed in exactly the same manner
as the observations.

Images were generated by combining the M87 observations after
subtracting background and correcting for exposure (which included all
the effects mentioned above). When generating merged images, to
exclude the readout stripe from the bright nucleus, we excluded
rectangular regions extending from the M87 nucleus to the edge of the
ACIS chip for each pointing (since the orientation of the readout
stripe varies from pointing to pointing). The backgrounds were
normalized by time and small corrections were made using the counts in
the 10-12 keV energy band in each image and corresponding background
file. Exposure maps in each energy band were computed separately and
included vignetting and detector response appropriate to the observed
spectrum, assumed to be emission from thermal gas with the redshift,
galactic absorption and elemental abundance of the Virgo cluster. For
quantitative imaging and spectral analysis we relied on the more
recent ACIS-I observations alone.

\begin{figure} 
\centerline{\includegraphics[width=0.85\linewidth]{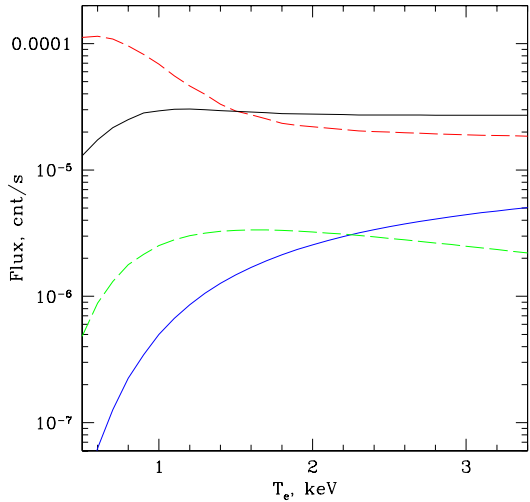}}
  \caption{The emissivity of  an
  optically thin plasma as a function of temperature as observed by
  the Chandra ACIS-I detector (in units of counts s$^{-1}$ with
  arbitrary normalization). For the 0.5-1.0 keV soft band (upper
  dashed line), the emissivity strongly increases for temperatures
  below 1~keV. For the 1.2-2.5 keV band (upper solid curve), the emissivity
  is nearly independent of gas temperature for temperatures above
  $\sim 0.75$ keV. Hence, surface brightness maps in this band are
  maps of $\int n^2 dl$ where $n$ is the electron density and $l$ is
  the path length.  In the 3.5-7.5 keV energy band (lower solid
  curve), the photon flux per unit volume $F$, can be expressed as $F
  \propto p^2 \epsilon(T)/T^2$ where $\epsilon (T)$ is the gas
  emissivity, $p$ is the pressure, and $T$ is the temperature (see
  text for details). In the hard 3.5-7.5 keV energy band,
  $\epsilon(T)/T^2$ (lower dashed curve) depends only weakly on $T$
  (for temperatures from 1-3 keV), and hence, the 3.5-7.5 keV band
  image is approximately an image of the square of the pressure
  (projected along the line of sight), i.e., total photon flux
  $\propto \int p^2 dl$. For this figure, we assumed an abundance of
  0.75 of solar. Note that for abundances in the range of 0-1 of
  solar, only the soft band curve (upper dashed line) changes
  qualitatively. It shows no peak at low temperatures. }
\label{fig:bands}
\end{figure}

\begin{figure*} 
  \centerline{\includegraphics[width=0.51\linewidth]{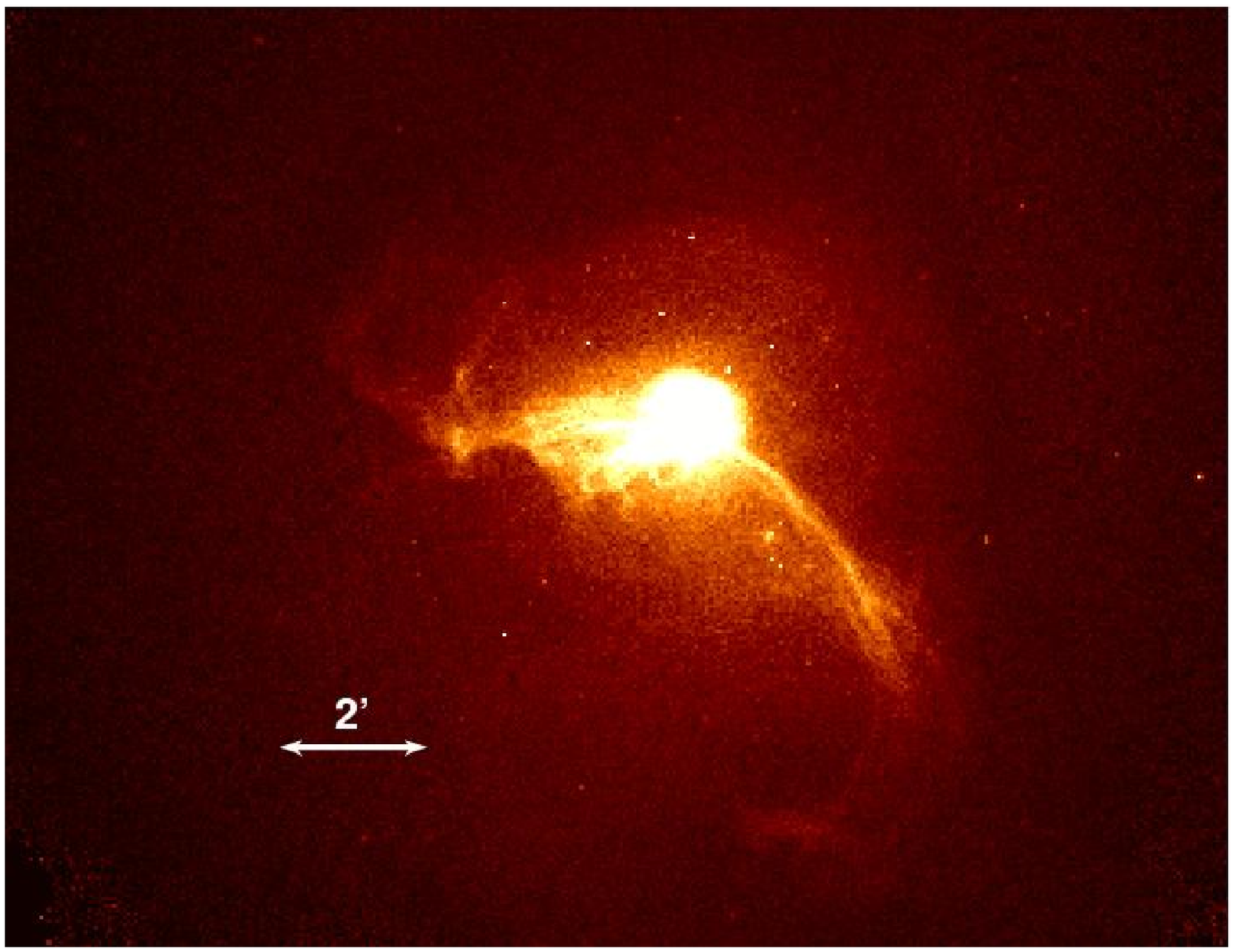}
  \includegraphics[width=0.45\linewidth]{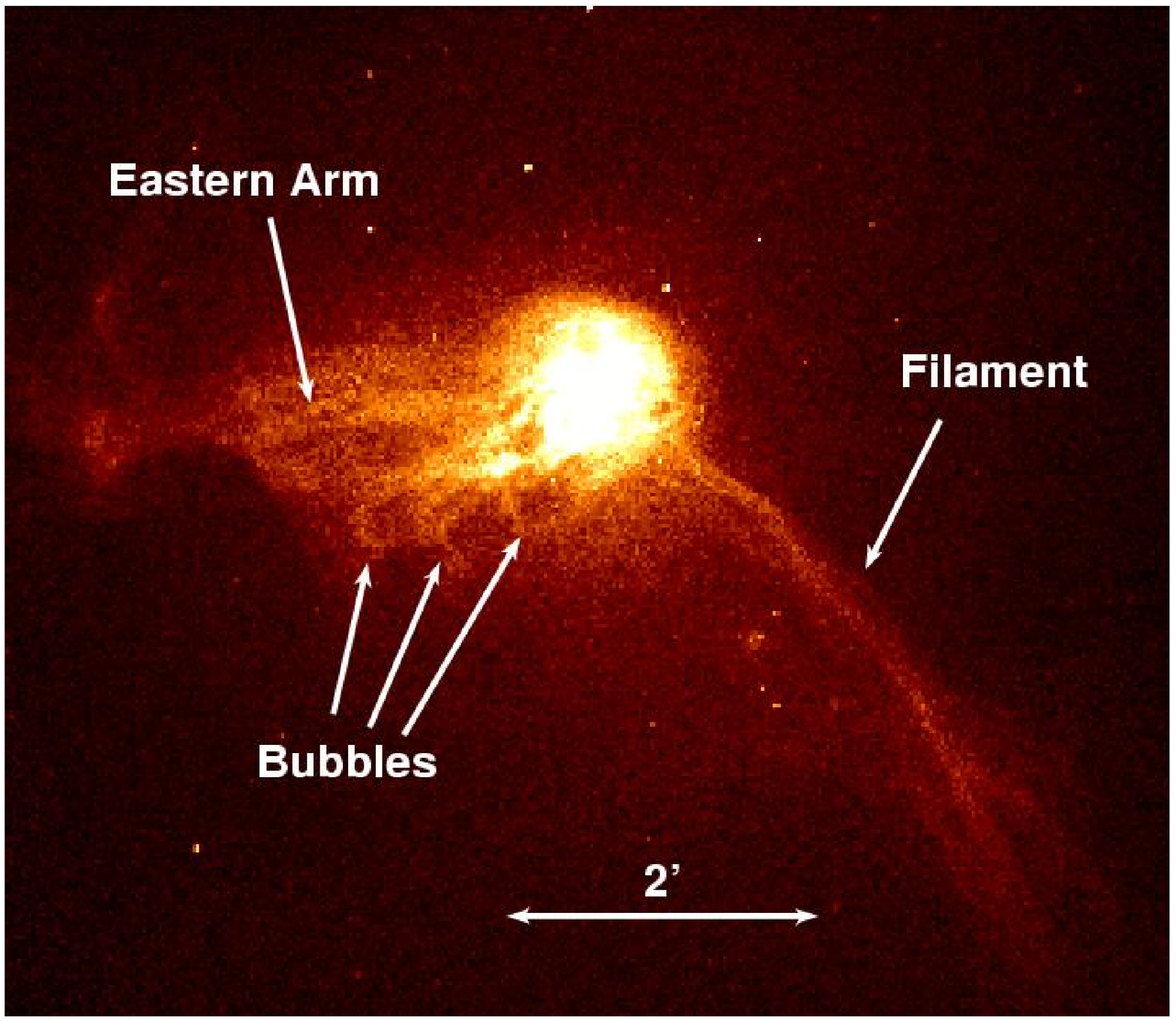}}
  \caption{Images of the 9 ACIS-I pointings after background
  subtraction and ``flat fielding'' in the energy band 0.5-1.0 keV
  with 1 pix=$3''$ at two different scales. a) The image at left shows
  the very prominent eastern and southwestern arms (see also
  Fig.~\ref{fig:overview}a, b where the are labeled in both the X-ray
  and radio images) b) The image at right shows a more detailed view
  of the tracery of filaments and suggests the similarity between the
  structures in the eastern and southwestern arms.  The long
  southwestern arm appears to be composed of several intertwined
  filaments (see location indicated by the label ``Filament''. The
  eastern arm can be interpreted as a series of bubbles (several
  labeled in right panel) at different evolutionary stages as they
  rise in the atmosphere of M87. The filamentary structures are very
  soft and are not apparent at energies above 2 keV (e.g., see
  Fig.~\ref{fig:hard}).  }
\label{fig:soft}
\end{figure*}

\begin{figure*}
\centerline{
  \includegraphics[width=0.5\linewidth]{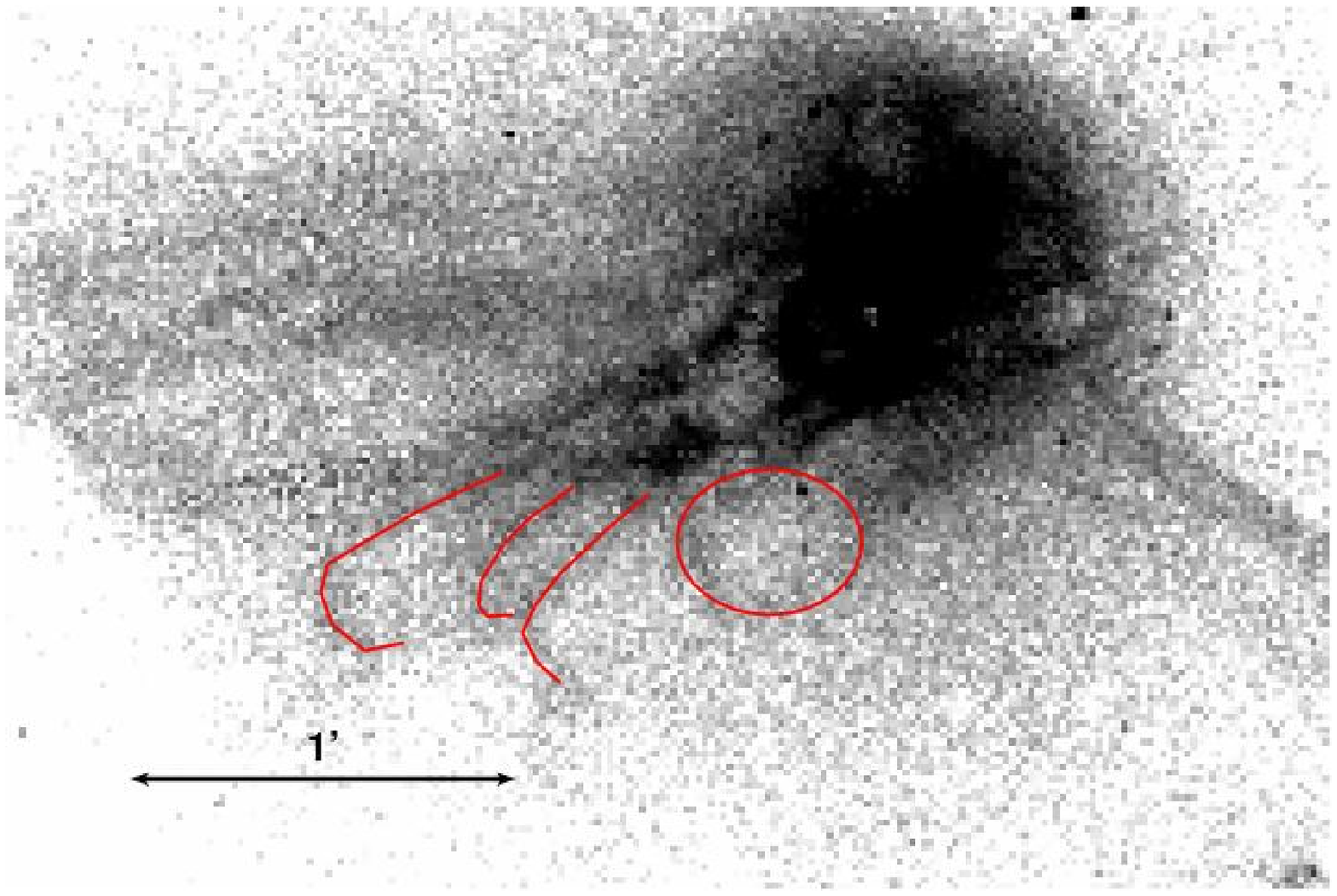}
  \includegraphics[width=0.45\linewidth]{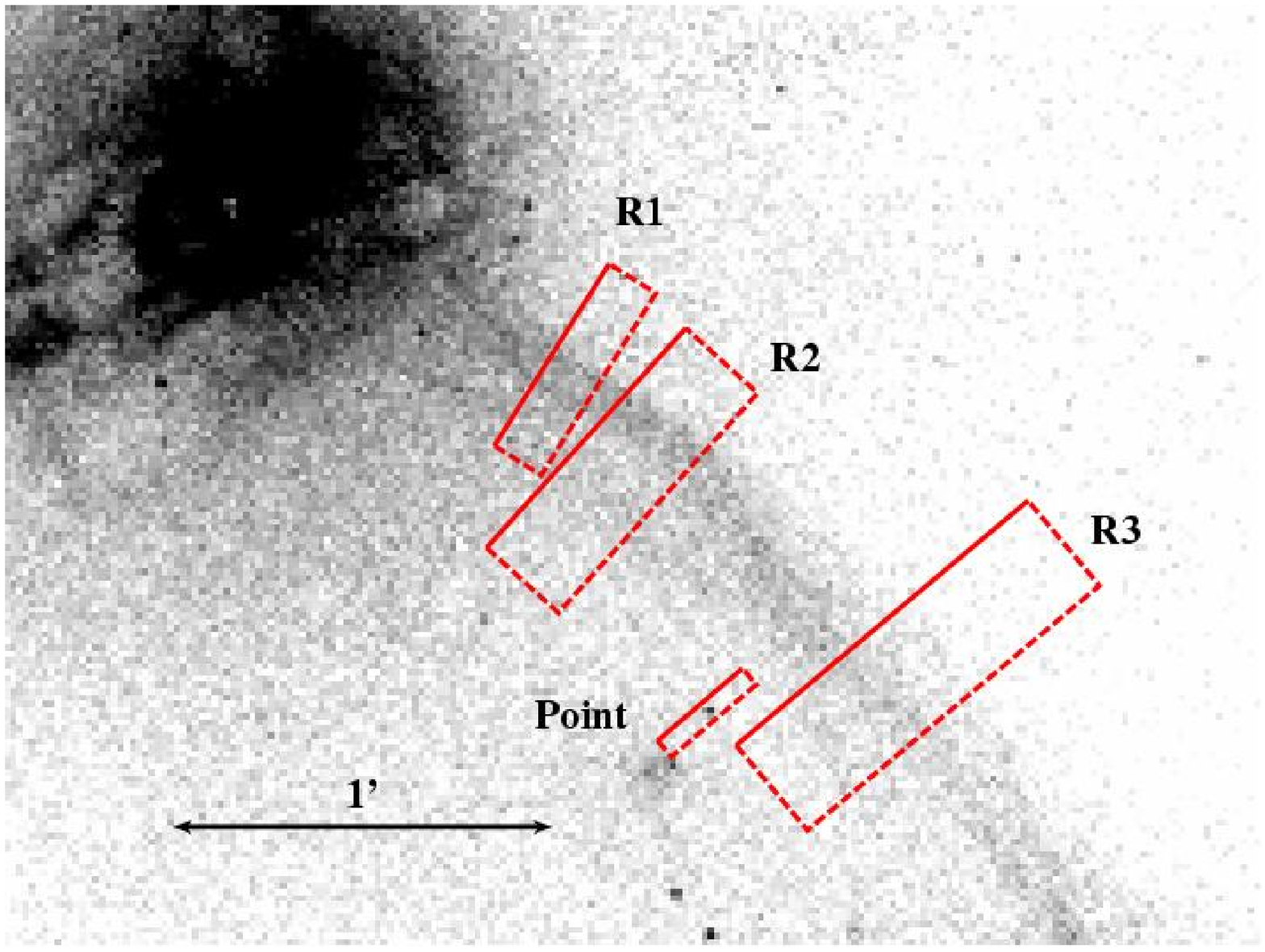}
}
  \caption{(a) The central region of the 0.5-1.0 keV band image
  showing the multiple buoyant bubbles rising in the M87 atmosphere. A
  series of four ``bubbles'' are outlined (see Fig.~\ref{fig:soft}b
  where the first three bubbles are indicated with arrows). (b) The
  0.5-1.0 keV image of a portion of the southwestern arm indicating
  the locations of projections and a comparison point source.}
\label{fig:filaments1}
\end{figure*}

To investigate the properties of the thermal gas that dominates the
Virgo core, we selected three energy bands that provide 1) an image of
the soft thermal gas (0.5-1.0 keV), 2) a density map (1.2-2.5 keV),
and 3) a pressure map (3.5-7.5 keV).  Using the full Chandra response,
we computed the predicted count rate for a unit volume of gas of unit
density for different energy bands as a function of gas temperature
(see Fig.~\ref{fig:bands}).  For the 0.5-1.0 keV band,
Fig.~\ref{fig:bands} (upper dashed line) shows that this energy band
(Fig.~\ref{fig:soft},~\ref{fig:filaments1}) focuses on the the soft
emission.

For the 1.2-2.5 keV band, Fig.~\ref{fig:bands} (upper solid line)
shows that the count rate for a unit volume of unit density is
independent of gas temperature, for gas temperatures above about
1~keV.  If we express
the count rate as

\begin{equation}
C \propto \int n_e^2 \epsilon(T) dl
\label{eq:fx}
\end{equation}

\noindent
where $n_e$ is the electron density, $\epsilon(T)$ is the volume
emissivity of the gas convolved with the Chandra response, and $l$ is
the path length along the line of sight, the independence of the
1.2-2.5 keV band emissivity on temperature implies that $C ({\rm
1.2-2.5~ keV}) \propto \int n_e^2 dl$. Thus, the 1.2-2.5 keV band
image (Fig.~\ref{fig:density}a, b) is a ``density'' image (actually
the square of the density integrated along the line of sight).

The properties of the 3.5-7.5 keV band can be determined by rewriting
eq.~\ref{eq:fx} using the ideal gas law, $p\propto n_e T$ to 
eliminate the gas density. We find

\begin{equation}
C \propto \int p^2 \epsilon(T)/T^2 dl
\label{eq:px}
\end{equation}

The lower dashed curve in Fig.~\ref{fig:bands} is $\epsilon(T)/T^2$
for a unit density and unit volume of gas.  For gas temperatures from
1 to 3 keV, Fig.~\ref{fig:bands} shows that this expression varies
only weakly with temperature.  Hence, eq.~\ref{eq:px} in the hard band
can be written as $C({\rm 3.5-7.5~ keV}) \propto \int p^2 dl$. Thus,
the hard band (Fig.~\ref{fig:hard}a, b) provides a ``pressure'' map for
the gas (actually the square of the pressure integrated along the line
of sight).  Hence, bright regions in the hard band indicate pressure
enhancements which may be characteristic of shocks.  In the analysis
and discussion below, we utilize these three energy bands to
investigate the gas around M87.

\subsection{Spitzer Observations}

\begin{figure*}
  \centerline{ \includegraphics[width=0.25\linewidth]{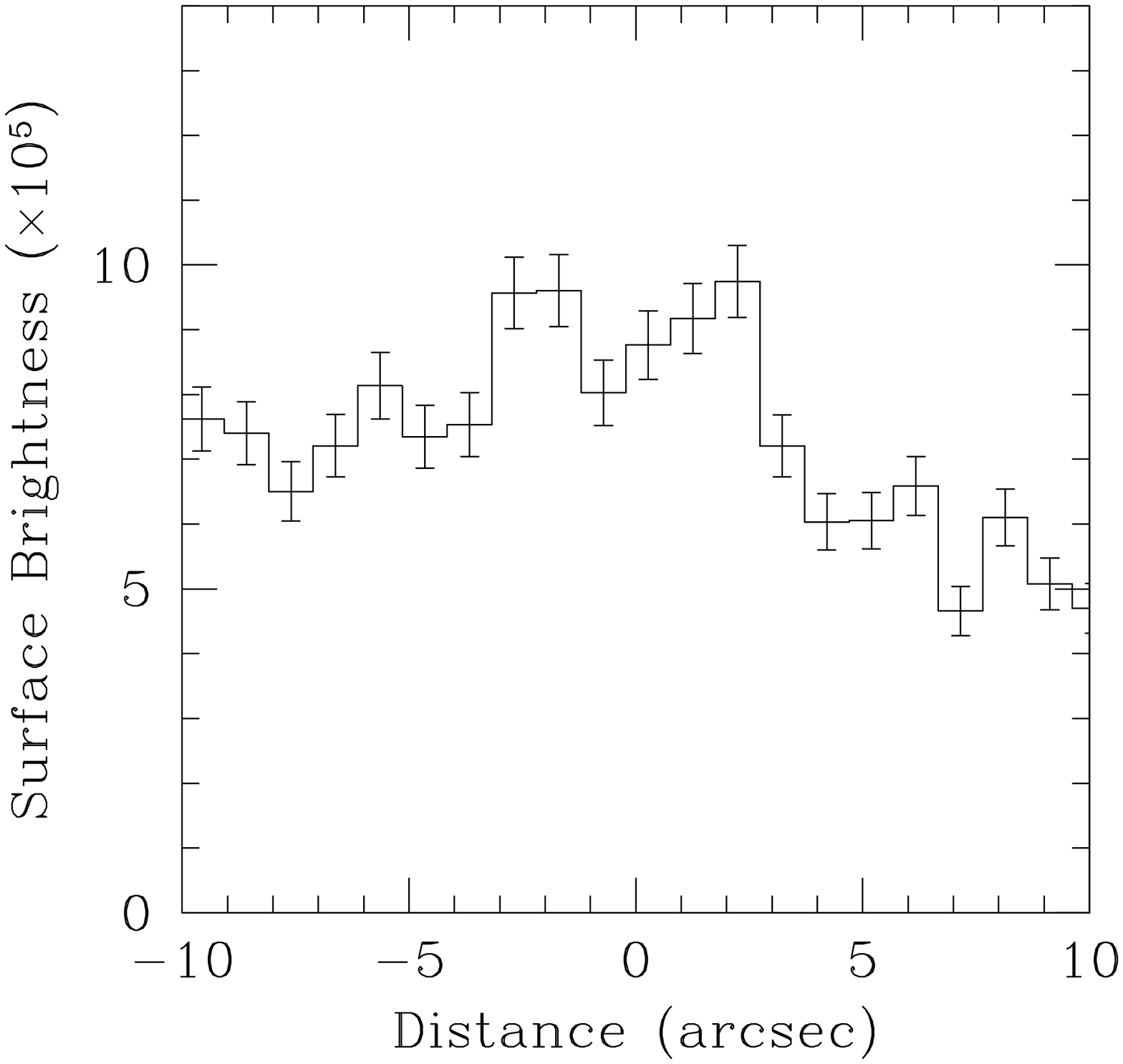}
    \includegraphics[width=0.25\linewidth]{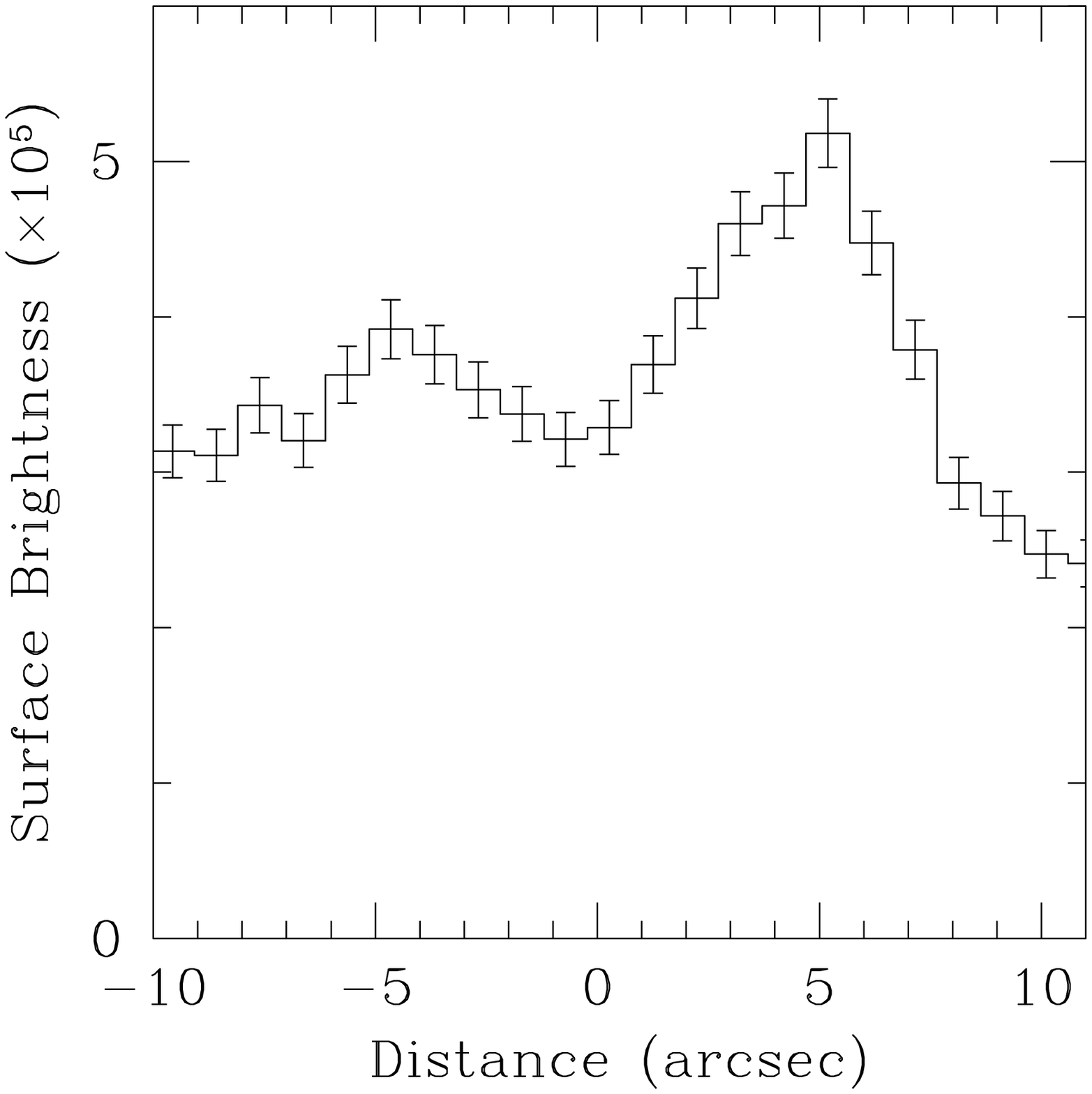}
    \includegraphics[width=0.25\linewidth]{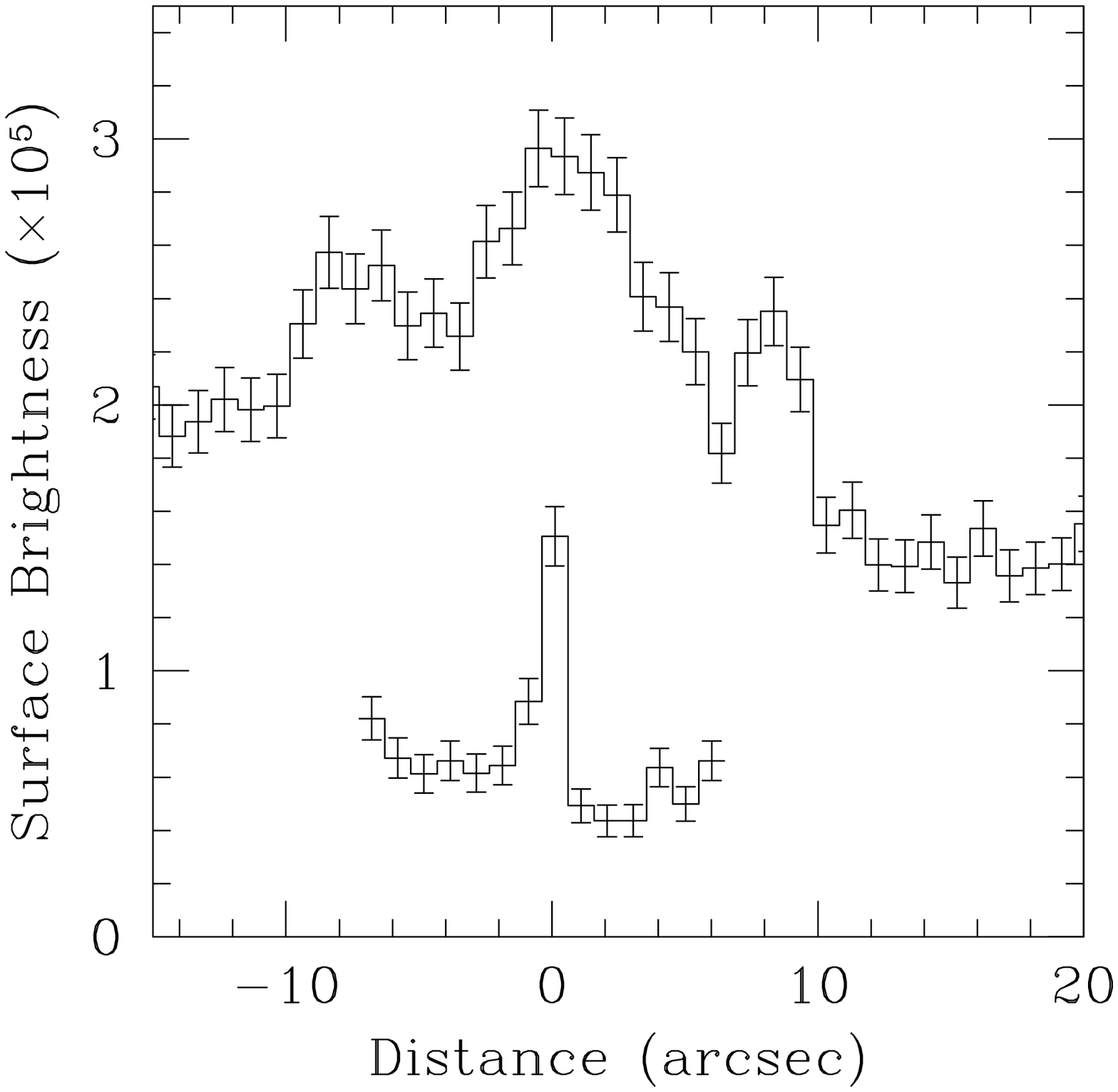}
    \includegraphics[width=0.25\linewidth]{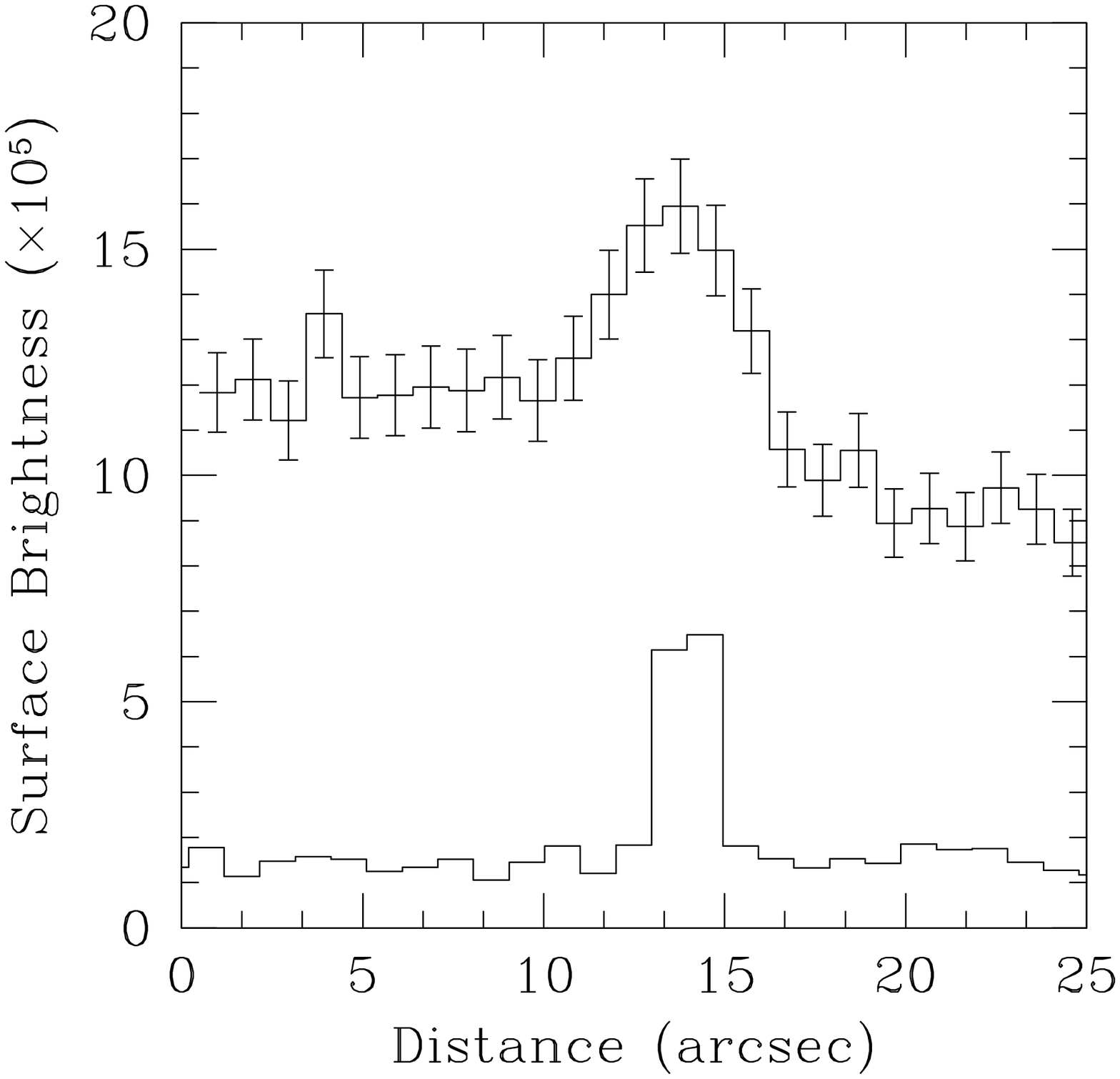} } \caption{The
    0.5-1.0 keV projections across the southwestern arm in the regions
    marked in Fig.~\ref{fig:filaments1}b. The figures show the complex
    structure along the arm.  a) The first panel (region R1 in
    Fig.~\ref{fig:filaments1}b) shows two peaks which may arise either
    from two separate, close, filaments or from the limb-brightened
    edges of a cylindrical filament. b) The second panel shows (region
    R2) a continuation of the southwestern arm consisting of
    two broad filaments. c)  The third panel (R3) shows three
    resolved structures (and a point source below, labeled ``Point''
    in Fig.~\ref{fig:filaments1}b). Comparison to
    Fig.~\ref{fig:filaments1}b shows that the westernmost filament
    appears to wrap around the central filament. d) The final panel shows
    a projection across a boundary of the outer bubble of the series
    in the eastern arm along with a scaled projection (lower
    histogram) of a point source $4'$ from the nucleus of M87, twice
    as far as the eastern arm filamentary region.  }
\label{fig:filaments2}
\label{fig:proj}
\end{figure*}

M87 also was imaged with the IRAC instrument (Fazio et al. 2004) on
the Spitzer Space Telescope on 2005 June 11 under Guest Observer
Program ID 3228.  The exposure time was 300 seconds for the central
region of the galaxy (\dataset[ads/sa.spitzer#0012673792]{Spitzer
dataset \# 0012673792}) but only 150 seconds for a larger region
surrounding the target (\dataset[ads/sa.spitzer#0010483200]{Spitzer
dataset \# 0010483200}).  To create the images in each waveband, the
Basic Calibration Data (BCD) were mosaiced using the Spitzer Science
Center software MOPEX (Makovoz et al. 2005).  Optical distortion was
removed in the process, and the images were subpixellated to
$0.86$~arcsec, which is a linear reduction by a factor of $\sqrt{2}$
in each dimension. Fig.~\ref{fig:core}c, the $4.5$\micron\ image
divided by an azimuthally symmetric $\beta$-model to remove the strong
stellar light gradient, shows the features associated with
the jet and counterjet. The synchrotron emission is visually clearest
at $\lambda = 4.5$\micron\ because this bandpass falls between the
starlight emission (strongest at $3.6$\micron\ and decreasing with
wavelength) and the strong PAH emission features (which can dominate
the $5.8$ and $8.0$\micron\ bands).

\section{ Imaging Results}

\begin{figure*} \centerline{
\includegraphics[width=0.95\linewidth]{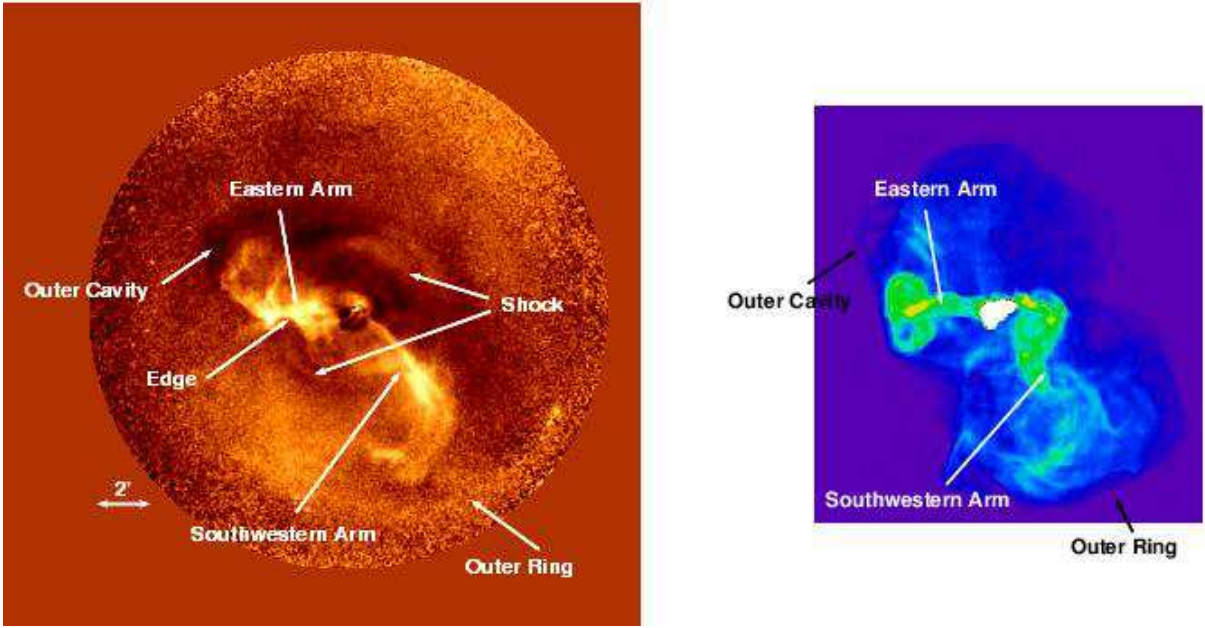} }
\caption{(a) The relative deviations of the surface brightness from a
radially averaged surface brightness model i.e., $[{\rm Data}-{\rm
Model}]/{\rm Model}$ over a broad energy band (0.5-2.5 keV).  The
shock, an outer cavity beyond the eastern arm, a sharp edge in the
eastern arm, and an outer partial ring are seen. We have excised the
prominent point sources from this image my substituting a local
background.  (b) The 90 cm VLA image from Owen et al. (1990) at the
same scale as the Chandra image shows the relationship between the
X-ray and radio structures.  In particular, the eastern and
southwestern arms are apparent in both X-ray and radio, the outer
X-ray cavity corresponds to an enhancement in the radio and the outer
ring (enhancement in X-ray image) lies just beyond the edge of the
large scale radio emission. The radio torus, at the end of the eastern
arm, is connected by the arm to the center of M87. The torus and arm
produces a ``mushroom'' shaped structure (cap and stem).}
\label{fig:overview}
\end{figure*}

Fig.~\ref{fig:soft}, the soft (0.5-1.0 keV) band image (see also
Fig.~\ref{fig:filaments1}), shows that the familiar soft X-ray
structures coincident with the radio arms (Fig.~\ref{fig:overview}a,
b) are dominated by a filamentary web that extends into a series of
bubbles to the south of the eastern arm (Fig.~\ref{fig:soft}b,
\ref{fig:filaments1}a). The hard band (3.5-7.5 keV) image,
Fig.~\ref{fig:hard}a, shows a shell with the characteristic pressure
signature of a weak shock.  In addition to these prominent features,
the images (see Fig.~\ref{fig:overview} for an overview of the large
scale structures) show:

\begin{itemize}
\item cavities at the end of the jet that correspond to radio emitting
  plasma (seen at 6 cm; see Fig.~\ref{fig:core}a labeled ``Counter Jet
  Cavity'' and ``Jet Cavity'' and compare to 6 cm image in
  Fig.~\ref{fig:core}b and $4.5$\micron\ Spitzer IRAC image in
  Fig.~\ref{fig:core}c). On the counter jet side, small scale
  structures in the IR and radio (Fig.~\ref{fig:core}b, c labeled with
  arrows) coincide. The counter jet structures probably lie beyond the
  end of the beamed jet.  Shi et al. (2006) found that several of the
  knots visible in the IRAC images, including the counter jet
  structures, have spectral indices $f_\nu \propto \nu^{-\alpha}$ with
  $\alpha$ ranging between $0.7$ and $0.9$.
\item a very clear demarcation of the counter jet cavity (marked
  ``Counter Jet Rim'' in Fig.~\ref{fig:core}a).
\item an over-pressurized egg-shaped region (most likely thermal gas
  since there is no corresponding non-thermal radio emission, e.g.,
  see Fig.~\ref{fig:core}b) of radius
  $\sim0.6'$ indicated by the lower arrow in Fig.~\ref{fig:hard}b and
  outer (cyan) contour in Fig.~~\ref{fig:density}b. This
  over-pressurized gas is being driven by the relativistic plasma in
  the radio cocoon (Fig.~\ref{fig:core}b and magenta contours in
  Fig.~~\ref{fig:density}b) that has been produced in the current
  outburst. 
\item a rim of soft X-ray emission (see shell labeled ``Cool Rim'' in
  Fig.~\ref{fig:density}b) surrounding the 
  overpressurized gas that represents gas swept up by the expanding
  ``piston'' (relativistic plasma of the cocoon).
\item a second weaker pressure ring at a radius of $\sim1'$ (indicated
  by the upper arrow in Fig.~\ref{fig:hard}b) may be the result of a
  secondary outburst, weaker than that which formed the main 13 kpc
  shock.
\item the very fine structure of the southwestern arm that may
  indicate a braided/twisted structure (see region marked ``Filament''
  in Fig.~\ref{fig:soft} and also
  Fig.~\ref{fig:filaments1}b). Although the arm itself is resolved,
  there appear to be finer filaments suggesting the importance of
  magnetic fields.
\item a sharp boundary at the southern edge of the eastern arm that
  does {\em not} correspond to any radio feature (marked ``Edge'' in
  Fig.~\ref{fig:overview}a). 
\item a new large X-ray cavity (labeled ``Outer Cavity'' in
  Fig.~\ref{fig:overview}a) at a radius of $4.8'-6.2'$ at an azimuth of
  $\theta=145$\dg~ lying just beyond the eastern X-ray arms (J2000,
  12:31:07.0, +12:27:0.0). The minimum energy required to inflate the cavity
  is $\sim10^{57}$ erg and its age (rise time) is about $7\times10^7$
  years. This feature is coincident with 90 cm radio emission (see 
  Fig.~\ref{fig:overview}b, the 90 cm radio image).
\item $\sim500$ point sources of which most are LMXBs and $\sim100$
  are background AGNs (e.g. some are seen in Fig.~\ref{fig:soft} and
  \ref{fig:core}a).
\end{itemize}

In the following discussion of the new X-ray observation of M87, the
intimate relationship between the thermal gas and the relativistic
radio emitting plasma should be kept in mind.  Fig.~\ref{fig:overview}
and ~\ref{fig:core} compare the X-ray and radio images and emphasize
this complex relationship.

\subsection{The Soft Filamentary Web}

The soft band image (Fig.~\ref{fig:soft}, \ref{fig:filaments1})
transforms our X-ray view of M87.  With the increased statistics of
the longer observation and the focus on the energy band below 1 keV,
we see a web of filamentary structures with length-to-width ratios of
up to $\sim50$.  This web is especially pronounced in the eastern arm.

The eastern arm appears to be produced by a series of buoyant
bubbles. The youngest is associated with a clearly visible ``bud'' in the
X-ray image (westernmost bubble labeled in Fig.~\ref{fig:soft}, marked
``Bud'' in Fig.~\ref{fig:core}a, and outlined circle in
Fig.~\ref{fig:filaments1}a; see also Forman et al. Fig.1b, c).  The
``bud'' has a radius of 0.9 kpc and is filled with radio plasma
(Fig.~\ref{fig:core}a and \ref{fig:core}b compare the X-ray and radio
images). The minimum energy required to inflate the ``bud'' is
$\sim10^{55}$ ergs for a spherical cavity (Forman et al. 2005).

Three bubbles, including the ``bud'', are marked ``Bubbles'' in
Fig.~\ref{fig:soft}b and four are highlighted in
Fig.~\ref{fig:filaments1}a.  We estimate the rise time of the four
features using the approximation
in Churazov et al. (2000; section 3.2) to derive the velocity,
$v_{rise}$ of buoyant bubbles. Equating the buoyancy force
to the drag force and taking the galaxy mass from Cot\'e et
al. (2001), we find $v_{rise}\approx400$ \kms~ over the whole inner 40
kpc of M87. With an average separation of the features of
approximately $0.5'$, the features differ in age by roughly
$6\times10^6$ years.

The series of bubbles, whose outer rims we only partially see in
Fig.~\ref{fig:filaments1}a, may continue into the eastern arm which
shows additional filamentary features.  These filaments may be
independent buoyant bubbles elongated by the rise of the large buoyant
bubble that produced the torus (cap of the column forming the eastern
radio arm seen and labeled in Fig.~\ref{fig:overview}b).

The X-ray filamentary structure is likely to be influenced by magnetic
filamentation or magnetic flux ropes. We know from Faraday rotation
data of the inner radio halo (Owen et al. 1990) and radio emission of
the large scale halo (Owen et al. 2000), that the magnetic fields are
highly filamented. The radio filaments are likely magnetic flux ropes.
The similarity in scale and geometry of the X-ray and radio filaments
argues for the importance of magnetic fields in their formation,
despite their different emission mechanisms.

The X-ray filaments that form the eastern arm are individually very
similar in width to the long, thin X-ray filaments that form the
southwestern arm (see Fig.~\ref{fig:soft}, \ref{fig:filaments1},
\ref{fig:overview}a).  Thus, while the eastern arm is significantly
broader than the southwestern arm, the structures which compose the
eastern arm are individually thin (see Fig.~\ref{fig:filaments1}a,
Fig.~\ref{fig:filaments2}).  We have used the marked regions in
Fig.~\ref{fig:filaments1}b to extract projections, shown in
Fig.~\ref{fig:filaments2}, across the southwestern arm and a point
source for comparison.  Fig.~\ref{fig:filaments2}a, the projection
closest to the center of M87 (R1 in Fig.~\ref{fig:filaments1}b) shows
a resolved arm (broader in extent than a point source) with two
subpeaks.  These could be either two separate filaments or the
limb-brightened edges of a cylindrical
tube. Fig.~\ref{fig:filaments2}b shows the onset of a second
filamentary structure about $10''$ east of the main
filament. Fig.~\ref{fig:filaments2}c shows the continuation of the
eastern filament, the main filament, and a third filament that in
Fig.~\ref{fig:filaments1} appears to be twisting around the central
filament. Comparison of the point source projections
(Fig.~\ref{fig:filaments2}c, d) to those of the filaments shows that
the filamentary structures have widths (FWHM) of about $4''$, 300 pc.

The integrated spectra of the arms are well characterized by a mean
temperature of $\sim1.5$ keV (see Belsole et al. 2001 and the
temperature map in Young et al. 2002).  However, Molendi (2002)
argued that the spectral data in the arms, on scales as small as
$4''\times4''$, require multiple components.  This is at least partly
caused by the blending of the thin filaments with intra-filament
emission. Also, it is likely that the filaments are composed of
still finer components that remain unresolved.  We will analyze the
detailed spectral properties of the arms, using the Chandra data, in a
later paper.

The Chandra observations highlight the differences seen in the
structure of the eastern and the southwestern arms.  While both arms
show long, narrow, spatially resolved filaments, the southwestern arm
has only a single set of filaments, while the eastern arm shows
multiple sets of filaments and bubbles that increase in scale with
distance from the center.  In addition, the radio plasma and X-ray
thermal plasma are co-spatial in the eastern arm while in the
southwestern arm, the X-ray and radio emitting plasmas avoid each
other with the X-ray plasma interior to the surrounding radio plasma
(see Fig.~\ref{fig:overview}).

\subsection{Weak Shocks}

\begin{figure*} 
\centerline{
\includegraphics[width=0.95\linewidth]{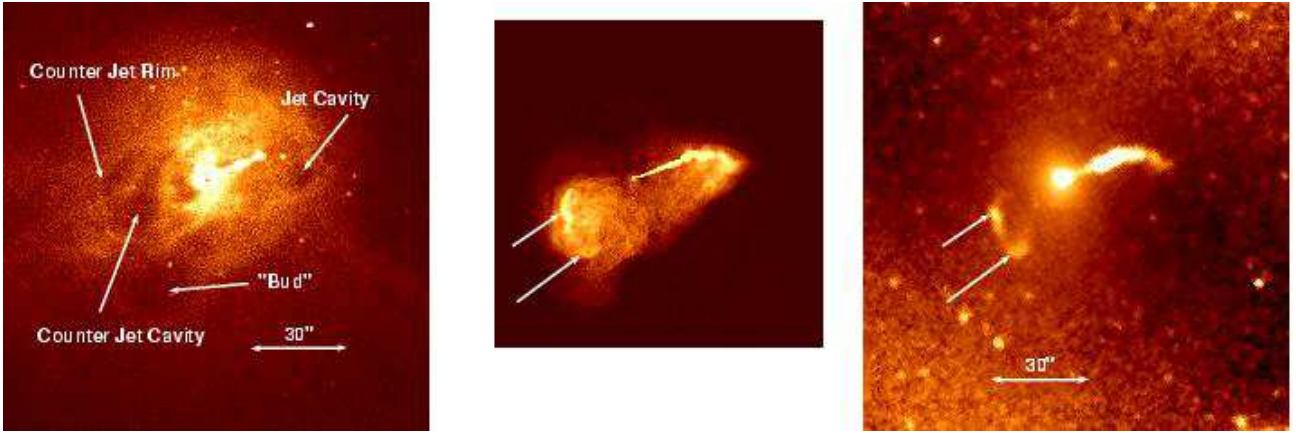} }
\caption{(a) The 0.5-2.5 keV band full resolution (1 pix = $0.492''$)
image of the entire data set after background subtraction and ``flat
fielding'' of the center of M87. b) The 6 cm VLA radio image from
Hines et al. (1989) showing the radio jet, and the synchrotron
emission from the cocoon.  The cocoon of relativistic plasma is the
``piston'' that mediates outbursts from the central SMBH and drives
shocks into the surrounding X-ray emitting, thermal gas. c) IRAC
$4.5$\micron\ image divided by a $\beta$ model to remove the strong
gradient of emission from the galaxy light.  Prominent X-ray features
of the central region show the counterjet cavity surrounded by a very
fine rim of gas and cavities to the west and southwest of the jet
after the jet passes the sonic point and the radio emitting plasma
bends clockwise. The innermost buoyant bubble (X-ray cavity, labeled
``bud'' in left panel) coincides with the radio synchrotron emission
extending south from the cocoon (center panel). The IRAC image shows
the emission from the nucleus and the jet. The IR jet emission ends
just before the feature ``Jet Cavity'' in the X-ray image. On the
counter jet side of the nucleus, two bright IR patches (labeled with
arrows in the IRAC image) lie within a ``C'' shaped region. The two
bright IR patches coincide with brighter regions of 6~cm emission (also
marked with arrows in the center panel) and associated with structures
$\theta$ and $\eta$ in Hines et al. (1989).  The IR emission (and the
coincident radio emission) lie at nearly 90\dg~ from the direction of
the jet (in projection) and arise from unbeamed emission.}
\label{fig:core}
\end{figure*}

\begin{figure*} 
 \centerline{\includegraphics[width=0.85\linewidth]{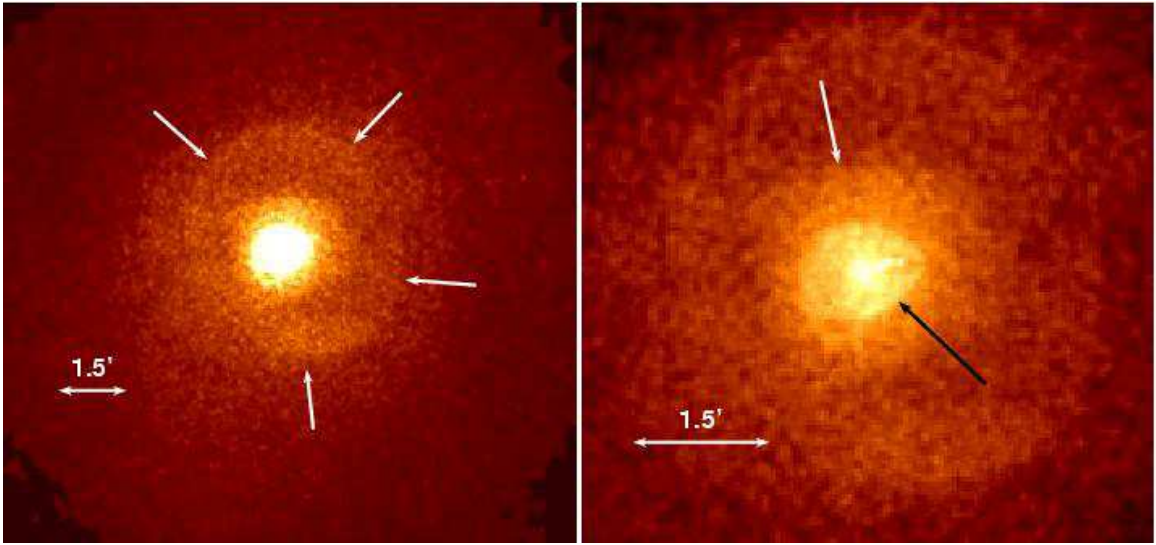}}
  \caption{ Images from the 3.5-7.5 keV energy band at different
  scales.  We have excised the prominent point sources from this image
  my substituting a local background. a) The 3.5 to 7.5 keV band image
  of the 9 ACIS-I pointings after background subtraction and ``flat
  fielding''.  The fine tracery seen in the soft band image,
  Fig.~\ref{fig:soft} is replaced by a nearly azimuthally symmetric
  ring (identified by arrows) of outer radius $\sim 2.75'$ (12.8
  kpc). As discussed in the text (see also Fig.~\ref{fig:bands}), this
  hard ring, a map of the square of the pressure projected along the
  line of sight, is the characteristic signature of a shock driven by
  an outburst from the central SMBH. b) The central region of the hard
  band image showing two additional regions of enhanced pressure. 
  An inner egg-shaped region of radius $\sim0.6'$ (outer edge marked
  with the lower arrow) with the narrow end of the ``egg'' aligned
  with the jet (barely visible in the image) was generated during the
  current AGN outburst from M87's SMBH. An earlier episode of activity
  is responsible for the 13 kpc shock. A second region of enhanced
  pressure has an outer edge of $\sim1'$ (marked with the upper arrow)
  may be a weak, secondary shock. } 
\label{fig:hard}
\end{figure*}

\begin{figure*} 
  \centerline{\includegraphics[width=0.45\linewidth]{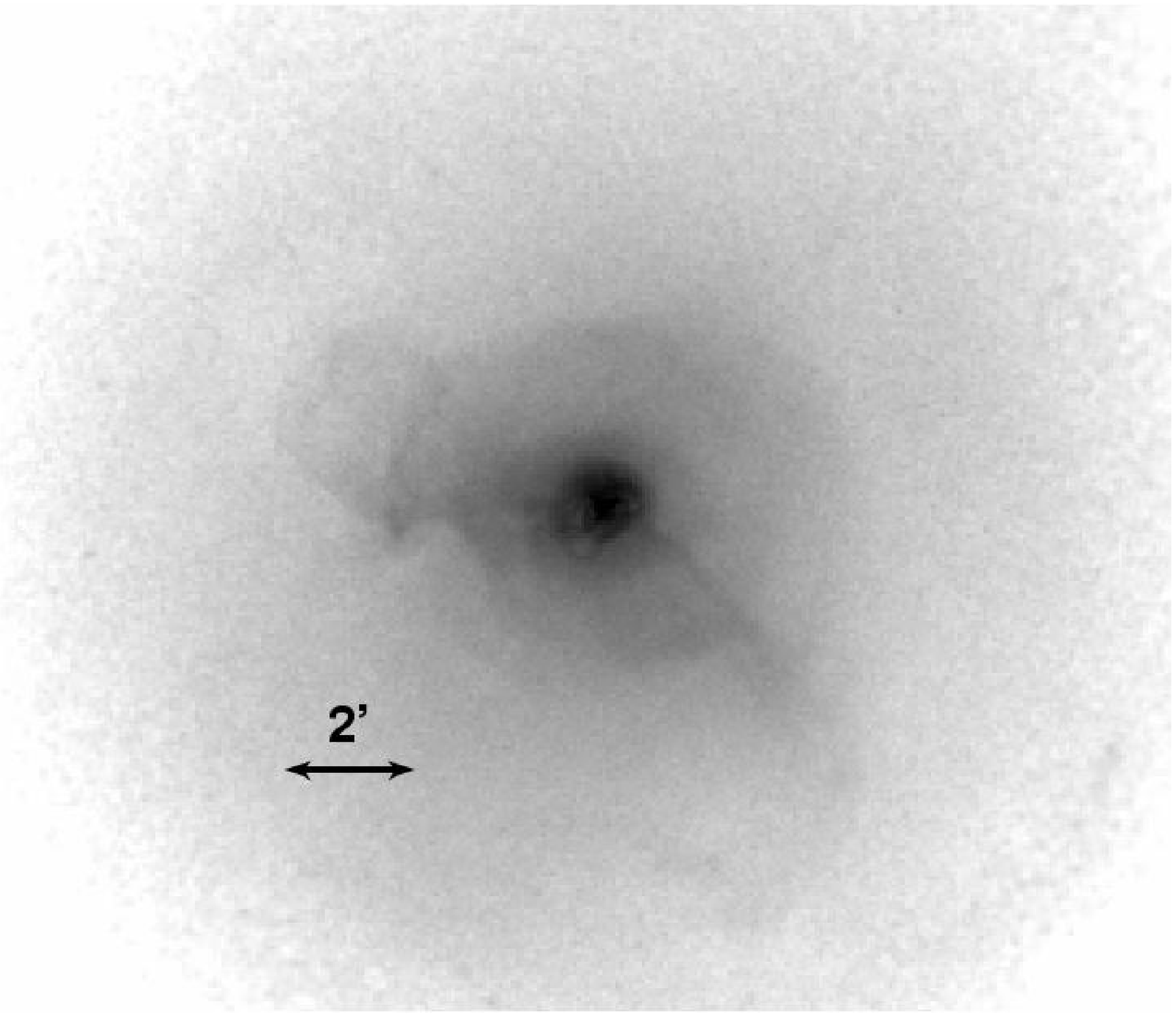}
\includegraphics[width=0.45\linewidth]{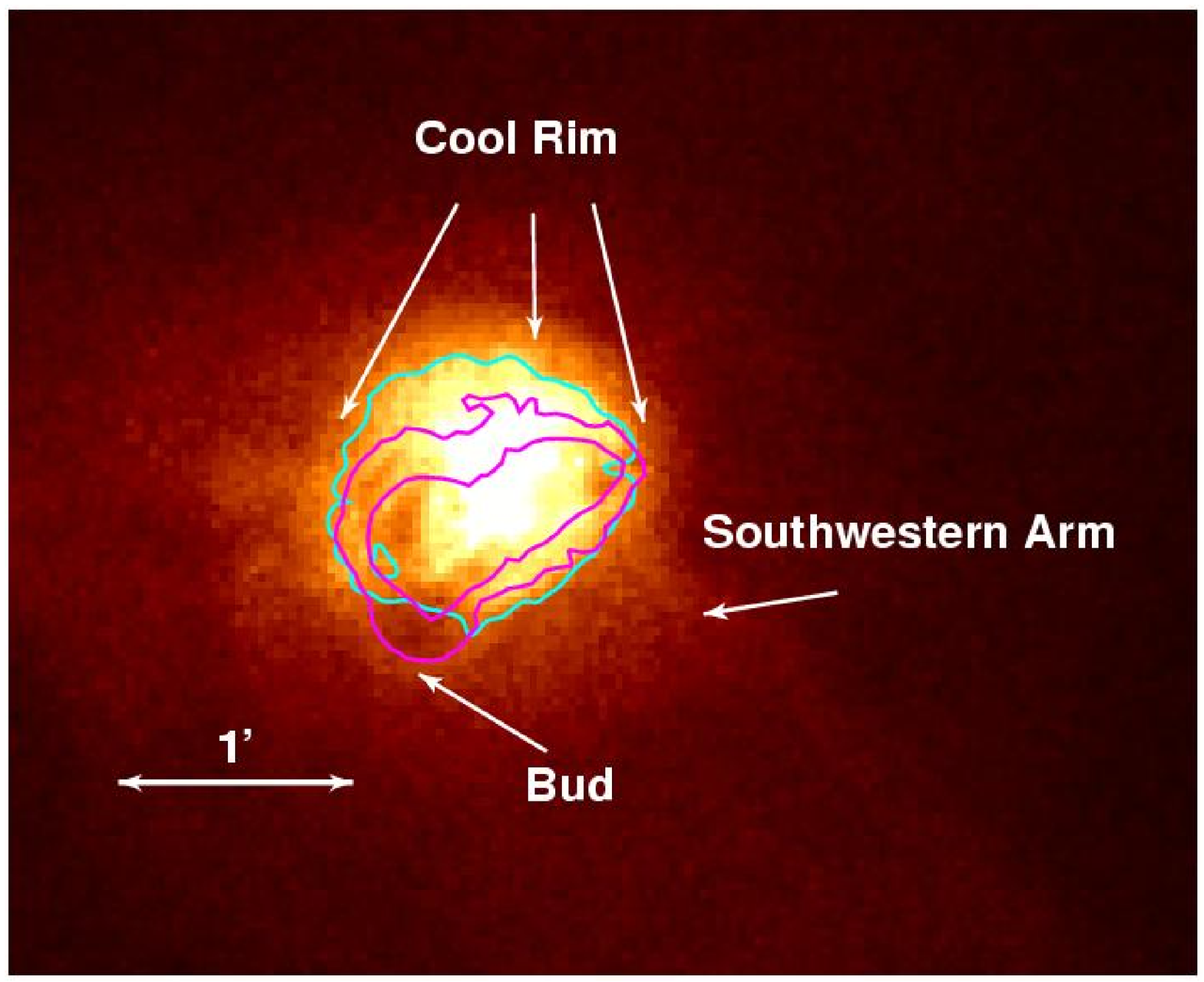} }
\caption{ a) Image from 1.2-2.5 keV derived from 9 ACIS-I pointings
after background subtraction and ``flat fielding''.  As noted in the
text (see also Fig.~\ref{fig:bands}), this image shows the the density
squared integrated along the line of sight. The pressure shock is seen
along with the two prominent arms. Many features, labeled in other
images are seen including the outer cavity, sharp edge, eastern and
southwestern arms labeled in Fig.~\ref{fig:overview}a, b and the
``bud'' labeled in Fig.~\ref{fig:core}. We have excised the prominent
point sources from this image my substituting a local background. b)
Central region of the 1.2-2.5 keV band image.  The two inner contours
(magenta) show the 6 cm VLA synchrotron emission which represents the
inner cocoon  (the ``piston'' of relativistic plasma that mediates
the energy output of the SMBH and drives the shocks into the
atmosphere of M87).  This outer contour is derived from the $0.6'$
high pressure region seen in Fig.~\ref{fig:hard}b and represents the
over-pressurized gas, that is being driven by the ``piston'' during
the current outburst. The figure shows the cool rim (labeled ``Cool
Rim'') that almost completely surrounds the over-pressurized region.
The cool rim is gas that has been displaced by the piston as it has
expanded into M87's atmosphere. Also marked are the bud (arrow denotes
its outer cool shell) and the base of the southwestern arm.}
\label{fig:density}
\end{figure*}

The hard band (3.5-7.5 keV) image, Fig.~\ref{fig:hard}a, is no less
striking than the soft band. We see, for the first time, the
unambiguous signature of a shock -- a nearly complete ring (indicated
by arrows in Fig.~\ref{fig:hard}a) of higher pressure gas. The outer
radius of the hard ring varies slightly as a function of azimuth,
$\theta$ (measured counter clockwise from west). From the southwest to
the northeast ($\theta = -45$\dg~ to 135\dg), over nearly 180\dg, the
outer radius is $R_{outer} = 2.85'$ (13 kpc).  In the eastern and
southeastern directions ($\theta=135$\dg-315\dg), the outer edge of
the ring is slightly smaller, $R_{outer}\sim2.5'$ (11.5 kpc).

Evidence for two additional features is seen in Fig.~\ref{fig:hard}b
at radii of $0.6'$ and $\sim1'$. The region of radius $0.6'$ (outer
boundary indicated by the lower arrow) is egg-shaped with the narrower
end aligned with the jet (barely visible in Fig.~\ref{fig:hard}b).
This region of high pressure was first reported by Young et al. (2002)
who noted a surface brightness ``front'' at a radius of $40''$. 
This over-pressurized region extends to the north, beyond the inner
bright radio cocoon seen in the 6 cm radio emission (see
Fig.~\ref{fig:core}b; Hines et al. 1989 and Fig.~\ref{fig:density}b).
We suggest that the plasma in the radio cocoon is the ``piston'' that
mediates the activity of the central supermassive black hole and
drives the pressure waves into the surrounding gas. The $0.6'$
over-pressurized region likely arises from the current episode of AGN
activity. In an earlier episode of activity, the ``piston'' drove the
13 kpc shock.  The second weak feature at $\sim1'$ (5 kpc), indicated
by the upper arrow in Fig.~\ref{fig:hard}b, may be the result of a
secondary outburst, weaker than that which produced the primary 13 kpc
ring.  Both the $0.6'$ and $1'$ features are seen in the surface
brightness profiles discussed below.

\begin{figure*}
  \includegraphics[width=0.450\linewidth]{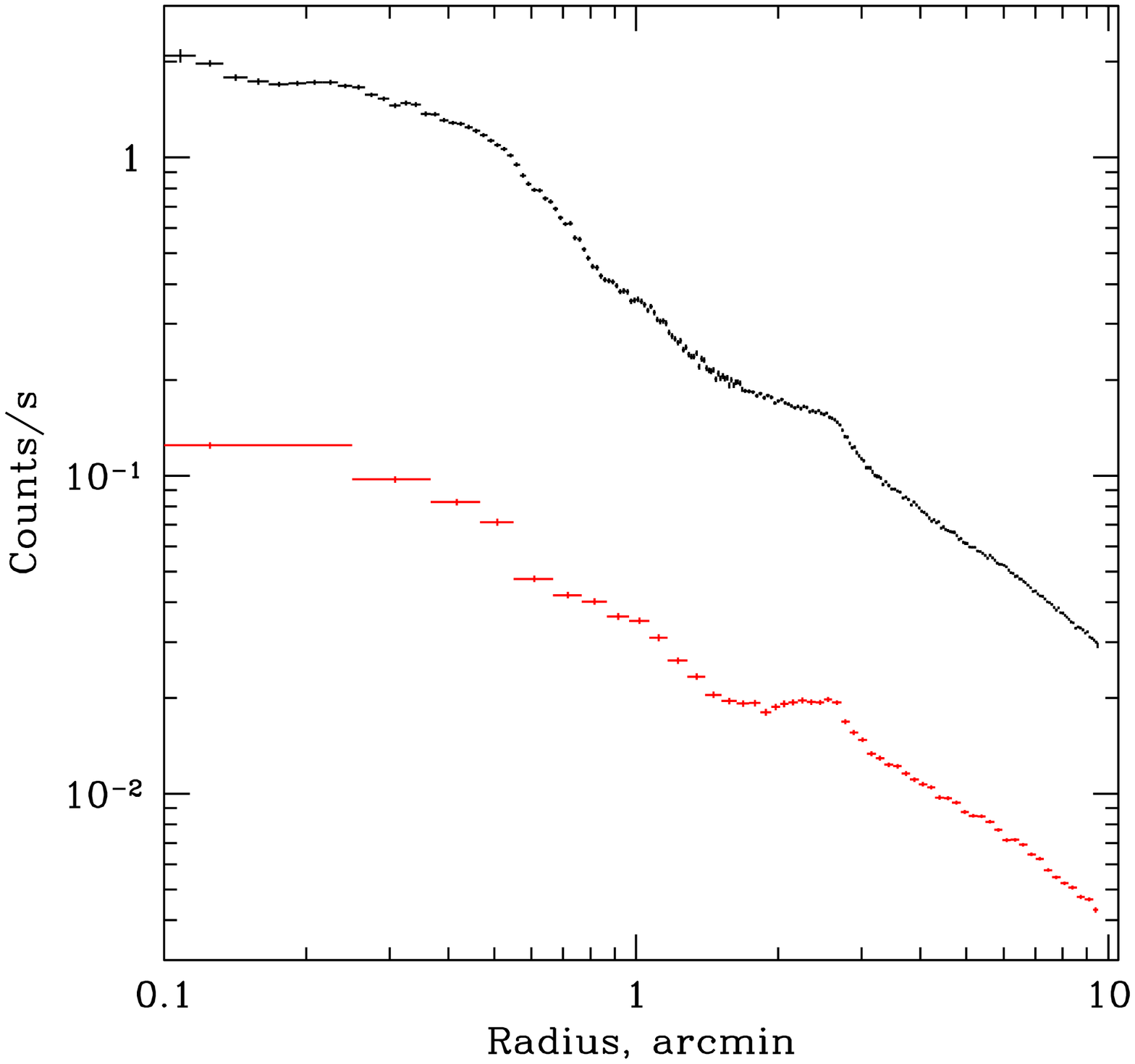}
  \includegraphics[width=0.450\linewidth]{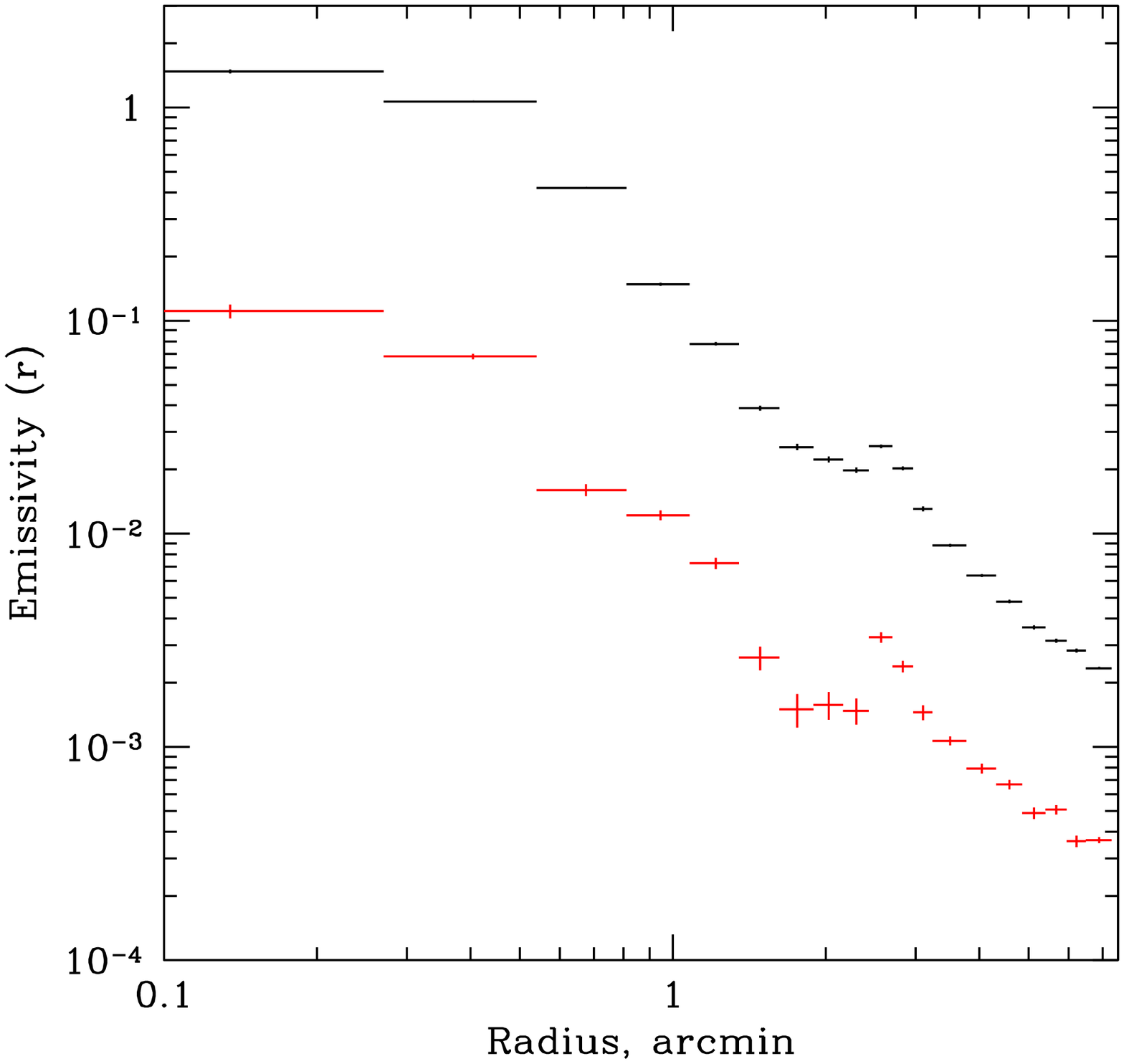} \caption{(a) The
  azimuthally averaged (sector of width 80\dg~ centered on north)
  surface brightness profiles in the 1.2-2.5 keV (upper curve) and
  3.5-7.5 keV (lower curve) bands show a sharp edge at $\sim0.6'$
  (most distinctly seen in the hard band as a decrease between the
  fourth and fifth data points), a moderate flattening of the profile
  at about $1'$, and a strong excess at $2-3'$. (b) The deprojected
  emissivity profiles, derived from the azimuthally averaged surface
  brightness profiles in the two energy bands 1.2-2.5 keV (upper
  curve) and 3.5-7.5 keV (lower curve). The deprojection shows the
  very pronounced feature at $2'-3'$ which represents the 13 kpc
  shock. In addition, for a spherical shock one expects a rarefaction
  region trailing the shock front (e.g. Zel'dovich \& Raizer 2002, p. 100)
  characterized by a density and temperature decrease below the
  upstream values. Such a rarefaction region is probably seen in
  Fig.~\ref{fig:sprofs}b at $r\sim1.5-2'$, where the emissivity shows a
  clear depression. The corresponding temperature decrease can also be
  identified in Fig.~\ref{fig:tprof}.  }
\label{fig:sprofs}
\end{figure*}

In our previous study, we detected only the surface brightness
enhancement associated with the main $13'$ shock, but lacked the
statistical precision to measure changes in gas temperature. We now
detect the increased pressure and spectral hardening characteristic of
a weak shock and can measure the temperature increase in the 13 kpc
shock.  Fig.~\ref{fig:sprofs}a shows the surface brightness profile in
the 1.2-2.5 keV (upper curve) and 3.5-7.5 keV (lower curve) energy
bands azimuthally averaged over an 80\dg~ wedge centered on north
where the surface brightness enhancement in the hard band lies at
nearly the same radial distance from the M87 nucleus. The most
pronounced feature is the shock between $2'$ and $3'$. Also seen in
the profiles is a decrease at $0.6'$ (most clearly seen in the hard
band between the fourth and fifth bins) which was noted above as being
over-pressurized gas produced by the present episode of activity (and
seen in the hard band image Fig.~\ref{fig:hard}b). The second
enhancement, possibly a weaker shock, is barely seen at $\sim1'$.  

To derive quantitative parameters of the main 13 kpc shock, we have
deprojected the surface brightness profiles.  In our analysis we
assume spherical symmetry but make no specific assumption about the
form of the underlying gravitational potential. We first calculate the
surface brightness (in a given energy band) in a set of annuli (or
wedges) and choose a corresponding set of spherical shells. The gas
parameters are assumed to be uniform inside each shell.  Outside $8'$
the emissivity was assumed to decrease with radius as a power law. The
projection can then be written as a convolution of the emissivities in
each shell with the projection matrix. The solution for emissivities
minimizing the $\chi^2$ deviation from the observed surface brightness
in the set of annuli can be easily found (see e.g. Churazov et al.
2003). The emissivities are then converted to electron densities using
the Chandra spectral response, evaluated for the spectrum with a given
temperature and abundance of heavy elements (see the discussion of
uncertainties introduced by the assumption of fixed temperature and
abundance at the end of this section). The resulting emissivity
profiles for the northern wedge are shown in Fig.~\ref{fig:sprofs}b.
We note that we are observing a disturbance in the gas that is only
approximately spherical. Hence, any shock features are broadened by
deviations from spherical symmetry.

\begin{figure}
\centerline{\includegraphics[width=0.850\linewidth]{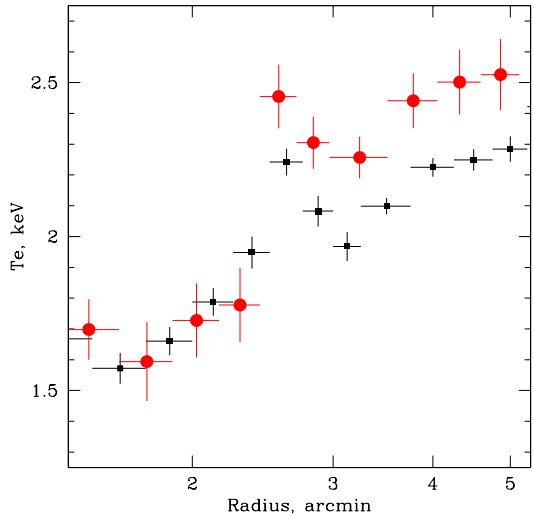}}
  \caption{The temperature profile derived from the 360\dg~
  azimuthally averaged data (small square symbols) and the wedge of
  width 80\dg~ (large round symbols) centered on north. We used the
  deprojected emissivity profiles in the 1.2-2.5 keV and 3.5-7.5 keV
  energy bands to derive the gas temperatures.  All telescope and
  detector effects have been included (as enumerated in the text for
  correcting the exposure maps). There is a clear variation in the
  temperature as a function of azimuth (the northern region appears
  hotter) which arises partly from the absence of a perfectly
  azimuthally symmetric distribution. The temperature rise associated
  with the shock is apparent in the temperatures derived for both the
  full 360\dg~ average and the northern wedge. We have not computed
  the temperatures at radii less than $1'$ due to the very complex
  structures seen at smaller radii (see Fig.~\ref{fig:core}).}
\label{fig:tprof}
\end{figure}

We have used the deprojected emissivity profiles in the two energy
bands to derive the gas temperature profile.  We applied the full
Chandra response with all telescope and detector effects included (as
enumerated above for correcting the exposure maps) to thermal gas
models. The resulting temperature profile for the northern wedge is
shown in Fig.~\ref{fig:tprof}. In addition to the surface brightness
profiles and deprojection (Fig.~\ref{fig:sprofs}a, b) for the northern
wedge, we repeated the analysis for the full 360\dg~ range of azimuths.
The results for both azimuthal ranges, shown in Fig.~\ref{fig:tprof},
demonstrate a marked rise in the gas temperature at $\sim2.5'$ radius.
We note that there is a variation in temperature as a function of
azimuth; the northern region (large round symbols) appears slightly
hotter than the mean (small square symbols).  As we discuss below,
despite these differences, the temperature jumps, and hence the
derived shock properties (using the Rankine-Hugoniot shock jump
conditions), are consistent.

To estimate the magnitude of the temperature rise, we fit the data
points in Fig.~\ref{fig:tprof} on either side of the shock. For the 360\dg~ averaged
profile, we consider three data points to be affected by the shock
(the peak and the one point on either side) while for the profile to
the north, we assume that only two points are affected (the peak and
the next point at larger radius; for the northern profile with less
data, the uncertainties are larger and the radial range of each bin is
larger).  The temperature rise is computed from a linear fit to the
temperature profile (using three points before the shock and
four points beyond) to the peak in each curve. For the two sets of
data, the 360\dg~ average and the northern wedge, we find $T_{shock} /
T_0 = 1.18\pm0.03$ and $1.24\pm0.06$ respectively ($T_0 = 1.90\pm0.04$
and $1.98\pm0.04$ keV; all uncertainties are statistical only).  The
temperature jumps, $T_{shock}/T_0$, are consistent and yield a Mach
number, $M\approx1.2$. We have verified that the deprojected hardness
ratios yield correct temperatures using direct spectral fits
(following the deprojection approach described in David et al. 2001).
However, these detailed temperature fits, although they agree with the
deprojection analysis results described above, have unacceptably large
values of $\chi^2$ and hence the resulting error bars are difficult to
interpret. We defer a detailed discussion of spectral fitting to a
later paper, but only note that the best fit values are in good
agreement with those derived from the hardness ratios. Within these
limitations, the temperature jump at the shock front is consistent
with expectations for the Mach 1.2 shock model of Forman et
al. (2005), using the Rankine-Hugoniot jump conditions for a monatomic
gas with $\gamma = 5/3$.

We have also analyzed the density jump at the 13 kpc shock to derive
an independent estimate of the shock Mach number.  Using the
azimuthally averaged, deprojected emissivity profile for the 1.2-2.5
keV band (Fig.~\ref{fig:sprofs}b), we derived the unperturbed density
distribution at the location of the shock by fitting a power law to
three points before and four after the shock.  The resulting density
jump is $\rho_{shock}/\rho_0 \approx 1.33\pm0.02$ which yields a Mach
number $M=1.22\pm0.02$ (statistical errors only). This independent
measurement of the shock strength from the density distribution agrees
with that derived from the temperature measurement.  {\em Thus, for
the shock at 13 kpc around M87, both density and temperature profiles
exhibit the properties of a classical shock in a gas with
$\gamma=5/3$.}  The age of the outburst that gave rise to the shock
must be approximately the radius of the shock divided the shock
velocity. This age, $t_{outburst} \sim R_{shock}/v_{shock} = 14$ Myr,
should slightly overestimate the age of the shock since the velocity
was higher in the past (see also Forman et al. 2005 who used a simple
model to derive an estimate of $t_{outburst} \sim 11$ Myr).

The gas temperature and densities derived above are subject to
systematic uncertainties. We mentioned above that projection and
asphericity introduce uncertainties. Projection effects will
undoubtedly reduce the temperature peak while any rarefaction region
behind the shock will increase it.  While strong shocks can
dramatically alter the gas properties behind the shock, the 13 kpc
shock in M87 is weak and, hence, behaves almost like an adiabatic wave
leaving little disturbance in the gas in the region behind the shock
that we use to derive the properties of the shock itself. Also,
abundance gradients (see Gastaldello \& Molendi 2002 and Matsushita et
al. 2003 for detailed discussions of the heavy element abundance
profiles in M87) can affect the radial profiles in different energy
bands. The effects on temperature determinations for the bands we have
selected are small. Over the temperature range from 1-3 keV, abundance
variations from 0 to 100\% of solar, change the temperature by at most
$\pm0.1$ keV.  We expect any abundance variations to occur gradually
as a function of radius (for our azimuthally averaged and deprojected
profiles). Furthermore, over the radial range from $1'-4'$ where we
are deriving the temperature and density properties of the shock, the
abundance gradients are small (see Matsushita et al. 2003).  Detailed
spectral fits, where the abundance and temperature are both free
parameters, confirm that any effects on the derived temperatures from
abundance variations are small.  For the gas density, the radial
profile would also vary only modestly.  For the observed range in
abundance in the radial range from $2'$ to $3'$ (e.g., Gastaldello \&
Molendi 2001), the change in the density would be less than 5\%. Thus,
uncertainties in the shock temperature and density jumps are modest.

\section{Conclusions}

M87 and its gaseous halo provide a unique laboratory for investigating
the interaction between a SMBH and the surrounding intracluster
medium.  Because of its low mean temperature of about 2.5 keV, the
energy range, 3.5-7.5 keV, where the pressure enhancements of modest
shocks are readily observed (Fig.~\ref{fig:hard}), lies within the
Chandra energy band.  Using the 500 ksec Chandra observation of M87,
we find:
\begin{itemize}
   \item a direct image of a weak shock, a region of enhanced
   pressure, at a radius of 13 kpc ((Fig.~\ref{fig:hard}a).
   \item  a central over-pressurized region (radius $0.6'$, seen
   in the hard band image Fig.~\ref{fig:hard}b, see also
   Fig.~\ref{fig:density}b) that surrounds the ``piston''(radio
   emitting relativistic plasma) responsible for mediating the SMBH
   activity.  The $0.6'$ over-pressurized region is itself surrounded
   by a cool rim of gas (Fig.~\ref{fig:density}b). During an earlier
   phase of activity, the ``piston'' drove a shock which we presently
   observe at 13 kpc. 
   \item at the shock, gas density and temperature jumps
   ($\rho_{shock}/\rho_0 \approx 1.33\pm0.02$ and $T_{shock}/T_0
   \approx 1.18\pm0.03$, respectively), yield consistent values of
   the shock Mach number, $M\approx1.2$, characteristic of a classical
   shock in a gas with $\gamma=5/3$.
   \item a web of soft filaments that may arise from a series of
   buoyant bubbles produced by small outbursts from the supermassive
   black hole at the center of M87. The filaments, the rims of the
   buoyant bubbles, are resolved by Chandra and have widths of
   approximately 300 pc,  although still finer, unresolved structures
   are likely to exist.
\end{itemize}

M87 shows a remarkably ``organized'' and coherent structure in the
eastern arm characteristic of a rising buoyant bubble transformed into
a torus in both the X-ray and radio images (see Owen et al. 2000 and
Churazov et al. 2001) and a second southwestern arm in both radio and
X-ray. However, although X-ray cavities are common in cooling core
clusters (e.g., Fabian et al. 2000, McNamara et al. 2000, Blanton et
al. 2001, Heinz et al. 2002, Mazzotta et al. 2003), the associated
thermal gas and radio emission characteristic of a large buoyant
bubble (Owen et al. 2000, Churazov et al. 2001; ``mushroom shaped''
feature) are, so far, only seen in M87.

While progress has been made in understanding galaxy clusters through
both Chandra and XMM-Newton observations, questions remain. The nature
of the filaments and importance of magnetic fields in their topology
requires additional modeling. Differences between the isothermal
``shock'' in Perseus (Fabian et al. 2006) and the clear hardening of
the shock in M87 need to be understood and observations of additional
clusters are required to place the M87 and Perseus observations in a
larger context.

\acknowledgements

We acknowledge stimulating discussions with T.~Bastian. We thank the
anonymous referee for helpful and constructive comments which
significantly improved the paper. This work was supported by contracts
NAS8-38248, NAS8-01130, NAS8-03060, the Chandra Science Center, the
Smithsonian Institution, MPI f\"{u}r Astrophysik, and MPI f\"{u}r
Extraterrestrische Physik. M. Begelman acknowledges support from NSF
grant AST-0307502. M.A.P. acknowledges NASA/LTSA grant \# NAG5-10777.
This work is based partly on observations made with the Spitzer Space
Telescope, which is operated by the Jet Propulsion Laboratory,
California Institute of Technology under NASA contract 1407.

\clearpage


\clearpage

\end{document}